\documentclass[aps,prl,superscriptaddress,showpacs,floatfix,twocolumn]{revtex4}

\usepackage{graphicx}   

\begin{document}


\title{Measurement of Bottom versus Charm as a Function of Transverse 
Momentum with Electron-Hadron Correlations in $p$+$p$ Collisions at 
$\sqrt{s}=200$~GeV}

\newcommand{\abilene}{Abilene Christian University, Abilene, TX 79699, U.S.}
\newcommand{\acadsin}{Institute of Physics, Academia Sinica, Taipei 11529, Taiwan}
\newcommand{\banaras}{Department of Physics, Banaras Hindu University, Varanasi 221005, India}
\newcommand{\barc}{Bhabha Atomic Research Centre, Bombay 400 085, India}
\newcommand{\bnlcoll}{Collider-Accelerator Department, Brookhaven National Laboratory, Upton, NY 11973-5000, U.S.}
\newcommand{\bnlphys}{Physics Department, Brookhaven National Laboratory, Upton, NY 11973-5000, U.S.}
\newcommand{\caucr}{University of California - Riverside, Riverside, CA 92521, U.S.}
\newcommand{\charlesczech}{Charles University, Ovocn\'{y} trh 5, Praha 1, 116 36, Prague, Czech Republic}
\newcommand{\ciae}{China Institute of Atomic Energy (CIAE), Beijing, People's Republic of China}
\newcommand{\cns}{Center for Nuclear Study, Graduate School of Science, University of Tokyo, 7-3-1 Hongo, Bunkyo, Tokyo 113-0033, Japan}
\newcommand{\colorado}{University of Colorado, Boulder, CO 80309, U.S.}
\newcommand{\columbia}{Columbia University, New York, NY 10027 and Nevis Laboratories, Irvington, NY 10533, U.S.}
\newcommand{\czechtech}{Czech Technical University, Zikova 4, 166 36 Prague 6, Czech Republic}
\newcommand{\dapnia}{Dapnia, CEA Saclay, F-91191, Gif-sur-Yvette, France}
\newcommand{\debrecen}{Debrecen University, H-4010 Debrecen, Egyetem t{\'e}r 1, Hungary}
\newcommand{\elte}{ELTE, E{\"o}tv{\"o}s Lor{\'a}nd University, H - 1117 Budapest, P{\'a}zm{\'a}ny P. s. 1/A, Hungary}
\newcommand{\fit}{Florida Institute of Technology, Melbourne, FL 32901, U.S.}
\newcommand{\fsu}{Florida State University, Tallahassee, FL 32306, U.S.}
\newcommand{\gsu}{Georgia State University, Atlanta, GA 30303, U.S.}
\newcommand{\hiroshima}{Hiroshima University, Kagamiyama, Higashi-Hiroshima 739-8526, Japan}
\newcommand{\ihepprot}{IHEP Protvino, State Research Center of Russian Federation, Institute for High Energy Physics, Protvino, 142281, Russia}
\newcommand{\illuiuc}{University of Illinois at Urbana-Champaign, Urbana, IL 61801, U.S.}
\newcommand{\instpasczech}{Institute of Physics, Academy of Sciences of the Czech Republic, Na Slovance 2, 182 21 Prague 8, Czech Republic}
\newcommand{\isu}{Iowa State University, Ames, IA 50011, U.S.}
\newcommand{\jinrdubna}{Joint Institute for Nuclear Research, 141980 Dubna, Moscow Region, Russia}
\newcommand{\kek}{KEK, High Energy Accelerator Research Organization, Tsukuba, Ibaraki 305-0801, Japan}
\newcommand{\kfki}{KFKI Research Institute for Particle and Nuclear Physics of the Hungarian Academy of Sciences (MTA KFKI RMKI), H-1525 Budapest 114, POBox 49, Budapest, Hungary}
\newcommand{\korea}{Korea University, Seoul, 136-701, Korea}
\newcommand{\kurchatov}{Russian Research Center ``Kurchatov Institute", Moscow, Russia}
\newcommand{\kyoto}{Kyoto University, Kyoto 606-8502, Japan}
\newcommand{\labllr}{Laboratoire Leprince-Ringuet, Ecole Polytechnique, CNRS-IN2P3, Route de Saclay, F-91128, Palaiseau, France}
\newcommand{\lawllnl}{Lawrence Livermore National Laboratory, Livermore, CA 94550, U.S.}
\newcommand{\losalamos}{Los Alamos National Laboratory, Los Alamos, NM 87545, U.S.}
\newcommand{\lpc}{LPC, Universit{\'e} Blaise Pascal, CNRS-IN2P3, Clermont-Fd, 63177 Aubiere Cedex, France}
\newcommand{\lund}{Department of Physics, Lund University, Box 118, SE-221 00 Lund, Sweden}
\newcommand{\mass}{Department of Physics, University of Massachusetts, Amherst, MA 01003-9337, U.S. }
\newcommand{\muenster}{Institut f\"ur Kernphysik, University of Muenster, D-48149 Muenster, Germany}
\newcommand{\muhlenberg}{Muhlenberg College, Allentown, PA 18104-5586, U.S.}
\newcommand{\myongji}{Myongji University, Yongin, Kyonggido 449-728, Korea}
\newcommand{\nagasaki}{Nagasaki Institute of Applied Science, Nagasaki-shi, Nagasaki 851-0193, Japan}
\newcommand{\newmex}{University of New Mexico, Albuquerque, NM 87131, U.S. }
\newcommand{\nmsu}{New Mexico State University, Las Cruces, NM 88003, U.S.}
\newcommand{\ornl}{Oak Ridge National Laboratory, Oak Ridge, TN 37831, U.S.}
\newcommand{\orsay}{IPN-Orsay, Universite Paris Sud, CNRS-IN2P3, BP1, F-91406, Orsay, France}
\newcommand{\peking}{Peking University, Beijing, People's Republic of China}
\newcommand{\pnpi}{PNPI, Petersburg Nuclear Physics Institute, Gatchina, Leningrad region, 188300, Russia}
\newcommand{\riken}{RIKEN Nishina Center for Accelerator-Based Science, Wako, Saitama 351-0198, JAPAN}
\newcommand{\rikjrbrc}{RIKEN BNL Research Center, Brookhaven National Laboratory, Upton, NY 11973-5000, U.S.}
\newcommand{\rikkyo}{Physics Department, Rikkyo University, 3-34-1 Nishi-Ikebukuro, Toshima, Tokyo 171-8501, Japan}
\newcommand{\saispbstu}{Saint Petersburg State Polytechnic University, St. Petersburg, Russia}
\newcommand{\saopaulo}{Universidade de S{\~a}o Paulo, Instituto de F\'{\i}sica, Caixa Postal 66318, S{\~a}o Paulo CEP05315-970, Brazil}
\newcommand{\seoulnat}{System Electronics Laboratory, Seoul National University, Seoul, Korea}
\newcommand{\stonybrkc}{Chemistry Department, Stony Brook University, Stony Brook, SUNY, NY 11794-3400, U.S.}
\newcommand{\stonycrkp}{Department of Physics and Astronomy, Stony Brook University, SUNY, Stony Brook, NY 11794, U.S.}
\newcommand{\subatech}{SUBATECH (Ecole des Mines de Nantes, CNRS-IN2P3, Universit{\'e} de Nantes) BP 20722 - 44307, Nantes, France}
\newcommand{\tenn}{University of Tennessee, Knoxville, TN 37996, U.S.}
\newcommand{\titech}{Department of Physics, Tokyo Institute of Technology, Oh-okayama, Meguro, Tokyo 152-8551, Japan}
\newcommand{\tsukuba}{Institute of Physics, University of Tsukuba, Tsukuba, Ibaraki 305, Japan}
\newcommand{\vandy}{Vanderbilt University, Nashville, TN 37235, U.S.}
\newcommand{\waseda}{Waseda University, Advanced Research Institute for Science and Engineering, 17 Kikui-cho, Shinjuku-ku, Tokyo 162-0044, Japan}
\newcommand{\weizmann}{Weizmann Institute, Rehovot 76100, Israel}
\newcommand{\yonsei}{Yonsei University, IPAP, Seoul 120-749, Korea}
\affiliation{\abilene}
\affiliation{\acadsin}
\affiliation{\banaras}
\affiliation{\barc}
\affiliation{\bnlcoll}
\affiliation{\bnlphys}
\affiliation{\caucr}
\affiliation{\charlesczech}
\affiliation{\ciae}
\affiliation{\cns}
\affiliation{\colorado}
\affiliation{\columbia}
\affiliation{\czechtech}
\affiliation{\dapnia}
\affiliation{\debrecen}
\affiliation{\elte}
\affiliation{\fit}
\affiliation{\fsu}
\affiliation{\gsu}
\affiliation{\hiroshima}
\affiliation{\ihepprot}
\affiliation{\illuiuc}
\affiliation{\instpasczech}
\affiliation{\isu}
\affiliation{\jinrdubna}
\affiliation{\kek}
\affiliation{\kfki}
\affiliation{\korea}
\affiliation{\kurchatov}
\affiliation{\kyoto}
\affiliation{\labllr}
\affiliation{\lawllnl}
\affiliation{\losalamos}
\affiliation{\lpc}
\affiliation{\lund}
\affiliation{\mass}
\affiliation{\muenster}
\affiliation{\muhlenberg}
\affiliation{\myongji}
\affiliation{\nagasaki}
\affiliation{\newmex}
\affiliation{\nmsu}
\affiliation{\ornl}
\affiliation{\orsay}
\affiliation{\peking}
\affiliation{\pnpi}
\affiliation{\riken}
\affiliation{\rikjrbrc}
\affiliation{\rikkyo}
\affiliation{\saispbstu}
\affiliation{\saopaulo}
\affiliation{\seoulnat}
\affiliation{\stonybrkc}
\affiliation{\stonycrkp}
\affiliation{\subatech}
\affiliation{\tenn}
\affiliation{\titech}
\affiliation{\tsukuba}
\affiliation{\vandy}
\affiliation{\waseda}
\affiliation{\weizmann}
\affiliation{\yonsei}
\author{A.~Adare} \affiliation{\colorado}
\author{S.~Afanasiev} \affiliation{\jinrdubna}
\author{C.~Aidala} \affiliation{\columbia} \affiliation{\mass}
\author{N.N.~Ajitanand} \affiliation{\stonybrkc}
\author{Y.~Akiba} \affiliation{\riken} \affiliation{\rikjrbrc}
\author{H.~Al-Bataineh} \affiliation{\nmsu}
\author{J.~Alexander} \affiliation{\stonybrkc}
\author{K.~Aoki} \affiliation{\kyoto} \affiliation{\riken}
\author{L.~Aphecetche} \affiliation{\subatech}
\author{R.~Armendariz} \affiliation{\nmsu}
% ...
\collaboration{PHENIX Collaboration} \noaffiliation

\date{\today}

\begin{abstract}

The momentum distribution of electrons from semi-leptonic decays of charm 
and bottom for mid-rapidity $|y|<0.35$ in $p$+$p$ collisions at $\sqrt{s} 
= 200$~GeV is measured by the PHENIX experiment at the Relativistic Heavy 
Ion Collider (RHIC) over the transverse momentum range $2<p_{{\rm 
T}}<7$~GeV/$c$.  The ratio of the yield of electrons from bottom to that 
from charm is presented. The ratio is determined using partial $D/\bar{D} 
\rightarrow e^{\pm} K^{\mp} X$~($K$ unidentified) reconstruction.  It is 
found that the yield of electrons from bottom becomes significant above 
4~GeV/$c$ in $p_{{\rm T}}$. A fixed-order-plus-next-to-leading-log~(FONLL) 
perturbative quantum chromodynamics (pQCD) calculation agrees with the 
data within the theoretical and experimental uncertainties. The extracted 
total bottom production cross section at this energy is 
$\sigma_{b\bar{b}}= 3.2 ^{+1.2}_{-1.1}({\rm stat}) {}^{+1.4}_{-1.3}({\rm sys}) 
\mu {\rm b}$.

\end{abstract}

        
\maketitle

Measurements of heavy flavor production (charm and bottom) in $p$+$p$ 
collisions present stringent tests for pQCD calculations.  For instance, 
while bottom production at the Tevatron is well described by 
next-to-leading order (NLO) pQCD~\cite{mateo1}, the cross section for 
charm production at high $p_T$, though compatible within the theoretical 
uncertainties, is higher than the preferred theoretical value by 
$\sim$50\%~\cite{cdf1}.  Measurement of heavy flavor in $p$+$p$ collisions 
also provides an important baseline for study of the medium created in 
relativistic heavy-ion collisions.  The PHENIX experiment at RHIC has 
measured single electrons from the semi-leptonic decay of heavy flavor at 
mid-rapidity in $p$+$p$ and Au+Au collisions at 
$\sqrt{s_{NN}}=200$~GeV~\cite{hq1,hq2}. Strong suppression of the single 
electron yield at high $p_T$, which includes contributions from both charm 
and bottom decays, was observed in central Au+Au collisions~\cite{hq2}. 
This effect is conventionally attributed to energy loss by the parent 
parton in the medium~\cite{jet1}; one also expects the energy loss 
suffered by bottom quarks to be significantly less than that suffered by 
charm quarks due to the difference in their masses~\cite{hq3,hq5}.  
Clearly, for both pQCD comparisons and the heavy-ion reference, one wants 
to disentangle the yields of charm and bottom at RHIC energies.

In this Letter, we present the the yield ratio of single electrons from 
bottom to those from heavy flavor at mid-rapidity in $p$+$p$ collisions at 
$\sqrt{s}\!=\!200$ GeV, using partial $D/\bar{D} \rightarrow e^{\pm} 
K^{\mp} X$~($K$ unidentified) reconstruction.

The data were collected with the PHENIX detector~\cite{ph1} in the 2005 and 
2006 RHIC runs using its two central arm spectrometers. Each spectrometer 
covers $|\eta|<$~0.35 in pseudorapidity and $\Delta \phi$~=~$\pi/2$ in 
azimuth. The arms include drift chambers~(DC) and pad chambers~(PC1,2,3) 
for charged particle tracking, a ring imaging $\check{\rm {C}}$erenkov 
detector~(RICH) and an electromagnetic calorimeter~(EMCal) for electron 
identification and triggering. Beam-beam counters~(BBCs), covering 
pseudorapidity 3.1~$<|\eta|<$ ~3.9, measure the position of the collision 
vertex along the beam ($z_{\rm vtx}$) and provide the interaction trigger. 
In the 2005 run, helium bags were placed in the space between the beam pipe 
and DC to reduce photon conversions. The bags were removed in 2006.

Two data sets are used for the analysis: (1) a minimum bias (MB) data set 
recorded with the BBC trigger, and (2) an electron enriched sample, 
recorded with a level-1 `ERT' trigger requiring a combination of EMCal 
and RICH information in coincidence with the BBC trigger. The BBC trigger 
cross section is 23.0~$\pm$~2.2 mb~\cite{ph2}. Since only $\sim$~53\% of 
inelastic $p$+$p$ collisions satisfy the BBC trigger condition, only a 
fraction of the inclusive electron production events are triggered. This 
fraction, which is $p_{\mathrm{T}}$ and process independent, is determined 
to be 0.79~$\pm$~0.02. After selection of good runs and a vertex cut of 
$|z_{\rm vtx}|<25$ cm, an integrated luminosity~($\int L dt$) in the ERT 
data of 1.77~pb$^{-1}$ in the 2005 run and 4.22~pb$^{-1}$ in the 2006 run 
are used for this analysis.

Charged particle tracks are reconstructed using the DC and PC1. The 
momentum resolution is $\sim$~1\% at $p_{{\rm T}} \sim$ 1~GeV/$c$, and the 
momentum scale is calibrated within 1\%. Electron identification (eID) is 
performed using the RICH and EMCal. The purity of the electron sample is 
better than 99\% for 1~$< p_{\rm{T}} <$~5 GeV/$c$~\cite{hq1}.
Our previous measurement~\cite{hq1} determined the spectrum of the single 
electrons from heavy flavor in the 2005 run.  Inclusive electron spectra 
from the 2005 run and the 2006 run are consistent within 5\% after taking 
into account a contribution from the increased photon conversion due to 
the absence of the helium bags.

The spectrum of the single electrons from heavy flavor is determined using 
the ``cocktail method''~\cite{hq1,hq2}. The electron spectrum from all 
known sources except semi-leptonic decay of heavy flavor is calculated 
using a Monte Carlo simulation and subtracted from the inclusive spectrum 
in the cocktail method. The dominant source of background is the $\pi^0$ 
Dalitz decay. The cocktail also includes contributions from 
quarkonium~($J/\psi$, $\Upsilon$) and the Drell-Yan process, which were 
neglected in our previous measurements~\cite{hq1,hq2}. These contributions 
are negligible~(smaller than 1\% in background) for $p_{\rm{T}}<1$~GeV/$c$ 
but, become significant at high $p_{\rm{T}}$~(above 10\% for 
$p_{\rm{T}}>2.5$~GeV/$c$)~\cite{morino}. The signal to background ratio 
increases with increasing $p_{\rm{T}}$, approaching unity for $p_{\rm{T}} 
\sim 3$~GeV/$c$~\cite{hq1}.

The systematic uncertainties of the inclusive electron spectrum includes 
the uncertainty in luminosity (9.6\%), geometrical acceptance (3\%), eID 
efficiency (2\%), and the ERT trigger efficiency (4\% at $p_{\rm{T}} > 
2$~GeV/$c$). The uncertainty in the cocktail method is $p_{\rm{T}}$ 
dependent (3\% at $p_{\rm{T}}$ $\sim$ 2~GeV/$c$, increasing to 9\% at 
9~GeV/$c$).

The ratio of $(b\rightarrow e)$ to $(c\rightarrow e+ b\rightarrow e)$ is 
extracted from the correlation between the heavy flavor electrons and 
associated hadrons~\cite{ua1}. The extraction is based on partial 
reconstruction of the $D/\bar{D} \rightarrow e^{\pm} K^{\mp} X$ decay. The 
invariant mass of unlike charge-sign electron-hadron pairs reveals a 
correlated signal below the $D$ meson mass of $\sim$1.9~GeV/$c^2$, because 
of the charge correlation in the $D$ decays. Pairs are formed between a 
trigger electron~($2.0\!<\!p_{\mathrm{T}}\!<\!7.0$~GeV/$c$) and an 
oppositely charged hadron~($0.4\!<\!p_{\mathrm{T}}\!<\!5.0$~GeV/$c$). The 
acceptances of positive and negative charged particles are forced to be 
identical by a geometrical acceptance cut. Since the momentum range of 
good charged kaon identification is limited, $K$ identification is not 
performed but the mass of all reconstructed hadrons is set to be that of 
the $K$. Most $e^{+}e^{-}$ pairs are then removed by an electron 
veto cut for the hadrons. The reconstructed mass of $e^{+}e^{-}$ pairs has 
a clear peak at low mass. The remaining background $e^{+}e^{-}$ pairs are 
removed by requiring $M_{ee}>80$~MeV/$c^2$, where the pair mass is 
calculated assuming both particles in the pair are electrons.

Depending on the origin of the trigger electrons, the inclusive 
reconstructed electron-hadron pairs are: (1) unlike-sign pairs from charm, 
(2) unlike-sign pairs from bottom, (3) combinatorial background where the 
electron is a background electron and (4) background from unlike-sign 
hadron-hadron pairs due to hadron contamination in the electrons. The main 
background source is the combinatorial background (3) and almost all 
background electrons are from $e^{+}e^{-}$ pair creation. Like-sign 
electron-hadron pairs are used to subtract this background. Since 
electrons from $e^{+}e^{-}$ pair creation and hadrons do not contribute to 
charge correlated signals, subtraction using like-sign pairs cancels out 
completely the combinatorial background where the trigger electron is from 
$e^{+}e^{-}$ pair creation~(3). Only the negligibly small~($<$1\%) 
contribution from $K^{0}_{e3}$ decay is not canceled out by the 
subtraction in the background~(3). The contribution from hadron 
contamination (4) is also less than a 1\% effect due to the excellent 
electron identification. After the subtraction, the reconstructed pairs 
include a contribution from bottom~(2) due to not identifying $K$. 
The contribution from bottom~(2) is much smaller than that from charm~(1) 
due to the bottom decay modes and kinematics. The reconstructed pairs also 
contain a signal from partial reconstruction of heavy flavor hadrons and a 
contribution from a combination of heavy flavor electrons and hadrons from 
jet fragmentation. The ratio of the yield of unlike-sign pairs to that of 
like-sign pairs is about 1.1 for invariant masses~($M_{eK}$) below 
1.9~GeV/$c^2$.

The fraction of bottom contribution to the electrons from heavy flavor 
is obtained as follows:
\begin{equation} \label{eq:bcr}
	\frac{(b\rightarrow e)}{(c\rightarrow e+ b\rightarrow e)} =
	\frac{\epsilon_c -\epsilon_{data}}{\epsilon_c-\epsilon_b},
\end{equation}
where $\epsilon_{data}$ is the tagging efficiency in real data and $\epsilon_{c(b)}$ is
the tagging efficiency for charm~(bottom) production.
These are defined as 
\begin{eqnarray}
	\label{eq:tagdata}
	\epsilon_{data} &\equiv &\frac{N_{pair}}{N_{e(HF)}} =
	\frac{N_{c\rightarrow tag}+N_{b\rightarrow tag}}
	     {c\rightarrow e + b\rightarrow e}, \\
	     \epsilon_{c} &\equiv& \frac{N_{c\rightarrow tag}}{c\rightarrow e} ,\quad \quad
	\epsilon_{b} \equiv \frac{N_{b\rightarrow tag}}{b\rightarrow e},
	\end{eqnarray}
where $N_{e(HF)}$ is the number of measured heavy flavor electrons. 
$N_{pair}$ is the number of background subtracted unlike-sign 
electron-hadron pairs for invariant mass within $0.4<M_{eK}<1.9$~GeV/$c^2$, 
which corresponds to the mass range of charmed hadrons. Here, 
$N_{c(b)\rightarrow tag}$ is the number of reconstructed signals within 
$0.4<M_{eK}<1.9$~GeV/$c^2$ for charm (bottom) production.

\begin{figure}[thb]
    \includegraphics[width=1\linewidth]{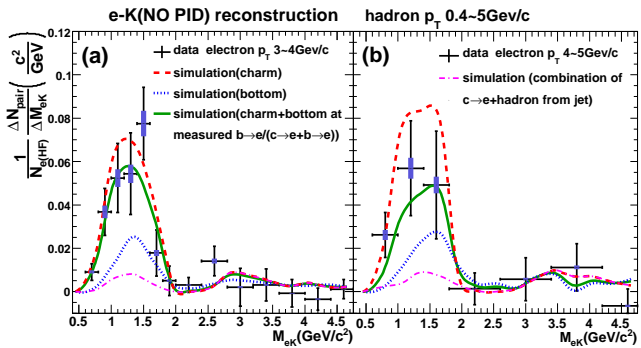}
\caption{\label{fig1} (color online) 
Comparison of data to a {\sc pythia} and {\sc evtgen} simulation of the 
invariant mass 
distributions in PHENIX acceptance for the reconstructed signal in the 2006 
run. The electron $p_{\mathrm{T}}$ range is 3.0 - 4.0~GeV/$c$~(a) and 4.0 - 
5.0~GeV/$c$~(b). The ratios, $(b\rightarrow e)/(c\rightarrow e+ 
b\rightarrow e)$, in solid lines are 0.26~(a) and 0.63~(b). Error 
bars~(boxes) indicate statistical~(systematic) uncertainties.
}
\end{figure}

Figure~\ref{fig1} shows the $M_{eK}$ distribution of the reconstructed 
signals, which is normalized by the yield of heavy flavor 
electrons~($N_{e(HF)}$) in the range $3\!<\!p_{\rm{ 
T}}\!<\!4$~GeV/$c$~(panel a) and $4\!<\!p_{\rm{ T}}\!<\!5$~GeV/$c$~(panel 
b). The tagging efficiency in real data, $\epsilon_{data}$, is determined 
by the integration of the $M_{eK}$ distribution in Fig.~\ref{fig1} from 
$M_{eK}=0.4$~to~1.9~GeV/$c^2$ as a function of electron $p_{\rm{T}}$.

The tagging efficiencies for charm and bottom production, $\epsilon_{c}$ 
and $\epsilon_{b}$, are calculated with the combination of {\sc pythia} 
and {\sc evtgen}~\cite{mot1,mot2}. {\sc pythia} is used to simulate charm 
and bottom production in $p$+$p$ collisions at $\sqrt{s}=200$~GeV and is 
tuned to reproduce heavy flavor hadron ratios: $D^+/D^0=0.45\pm 0.10$, 
$D_s/D_0=0.25\pm0.10$, $\Lambda_c/D^0=0.10\pm0.05$, $B^+/B^0=0.50$, 
$B_s/B_0=0.40\pm0.20$, and $B_{{\rm 
baryon}}$$/B^0=0.20\pm0.15$~\cite{morino,bib:cratio1,bib:cratio2,bib:cratio3,bib:PDG}. 
{\sc evtgen}, which is a Monte-Carlo simulation suited for decays of $D$ 
and $B$ hadrons, is used to simulate the semi-leptonic decays. The 
dashed~(dotted) lines in Fig.~\ref{fig1} show the $M_{eK}$ distributions 
of the reconstructed signal for the simulated charm~(bottom) production 
for an electron $3\!<\!p_{\rm{ T}}\!<\!4$~GeV/$c$~(panel a) and 
$4\!<\!p_{\rm{ T}}\!<\!5$~GeV/$c$~(panel b). Some fluctuations in the 
simulated curves in Fig.~\ref{fig1} come from the limited statistics in 
the simulation, but the statistical uncertainties in the simulation are 
negligible compared to that of the data. $\epsilon_{c(b)}$ is determined 
in the same way as $\epsilon_{data}$ from the $M_{eK}$ distribution for 
charm~(bottom) production. Since about $85\%$ of the extracted signal 
comes from partcial reconstruction of heavy flavor hadrons, the tagging 
efficiency is determined largely by decay kinematics and $\epsilon_{c(b)}$ 
can be determined with good precision. The dot-dash lines in 
Fig.~\ref{fig1} show the contribution from the combination of an electron 
from charm and hadrons from jet fragmentation for charm production. The 
solid line in Fig.~\ref{fig1} shows the sum of the $M_{eK}$ distributions 
for charm and bottom in the simulation with the ratio, $(b\rightarrow 
e)/(c\rightarrow e+ b\rightarrow e)$, obtained with Eq.~\ref{eq:bcr}.

Systematic uncertainties are categorized into two parts related to (1) 
$\epsilon_{data}$ in the real data analysis and (2) $\epsilon_{c}$ and 
$\epsilon_{b}$ in the simulation study. The dominant uncertainty in 
$\epsilon_{data}$ is the uncertainty in the number of heavy flavor 
electrons~($\sim$10\%). Uncertainty in $\epsilon_{data}$ also includes a 
background subtraction uncertainty~(1-10\%, $p_{\rm{ T}}$ dependent). 
Category (2) includes the uncertainties in geometrical acceptance~(3\%) and 
the event generator~($\sim$~8\% for charm and $\sim$~9\% for bottom). The 
event generator uncertainty is based on uncertainties, which are known in 
the production ratios of heavy flavor hadrons~($D^+/D^0$, $D_s/D^0$,etc.), 
known in the branching 
ratios~\cite{bib:cratio1,bib:cratio2,bib:cratio3,bib:PDG}, estimated 
in the momentum distribution of heavy flavor hadrons and 
estimated in the {\sc pythia} parameters.


\begin{figure}[htbp]
\includegraphics[width=0.85\linewidth]{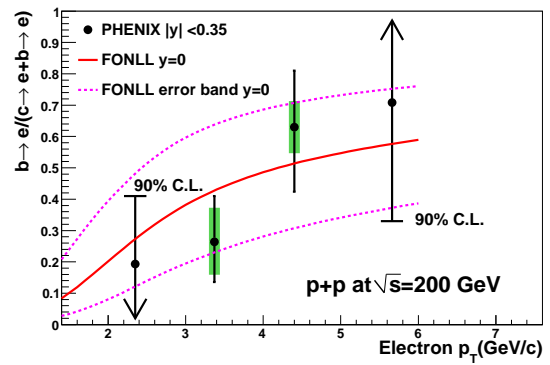}
\caption{\label{fig2} (color online) 
$(b\rightarrow e)/(c\rightarrow e+ b\rightarrow e)$ as a function of 
electron $p_{\mathrm{T}}$ compared to a FONLL calculation~\cite{mateo2}. 
The points show the experimental result. Vertical arrows are used to 
indicate upper and lower limits. The solid line is a FONLL prediction and 
the dotted lines represent the uncertainty of this FONLL prediction.
}
\end{figure}

Figure~\ref{fig2} shows the resulting bottom fraction, $(b\rightarrow 
e)/(c\rightarrow e+ b\rightarrow e)$ as a function of electron 
$p_{\mathrm{T}}$ compared to a fixed-order-plus-next-to-leading-log 
perturbative QCD calculation~(FONLL)~\cite{mateo2}. In this figure, the 
points show the measured $(b\rightarrow e)/(c\rightarrow e+ b\rightarrow 
e)$. For the bins with electron $p_{\rm{T}}$ ranges $2\!<\!p_{\rm{ 
T}}\!<\!3$ and $5\!<\!p_{\rm{ T}}\!<\!7$~GeV/$c$, 90\% C.L. and mean values 
are shown. The solid line shows the central value of the FONLL prediction 
and the dotted lines show its uncertainty.

\begin{figure}[tbh]
\includegraphics[width=0.95\linewidth]{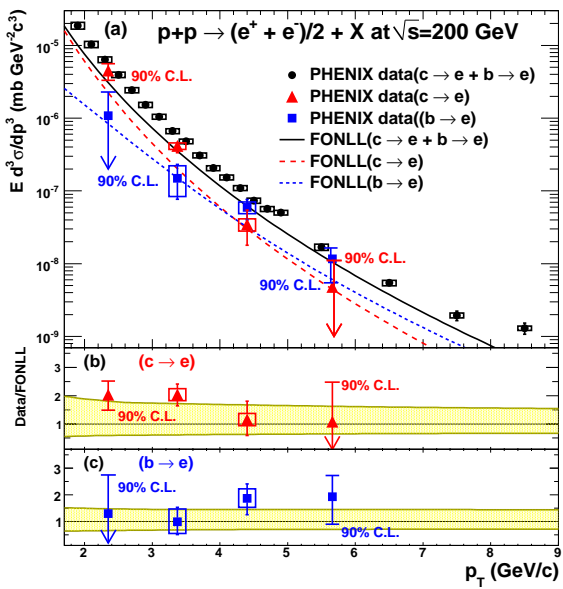}
\caption{\label{fig3} (color online) 
(a): Invariant cross sections of electrons from charm and bottom with the 
FONLL calculation~\cite{mateo2}. (b) and (c): The ratios of data points 
over the FONLL prediction as a function of electron $p_{\mathrm{T}}$ for 
charm and bottom. The shaded area shows the uncertainty in the FONLL 
prediction.
}
\end{figure}

In Fig.~\ref{fig3}, the single electron spectra for charm and bottom are 
measured from the ratio, $(b\rightarrow e)/(c\rightarrow e+b\rightarrow 
e)$, and the spectrum of the electrons from heavy flavor decays. The top 
panel shows the resulting single electron spectra from charm~(triangles) 
and bottom~(squares) compared to the FONLL 
predictions~\cite{mateo2}. The measured spectrum of single 
electrons~(circles) is also shown for reference. The middle~(bottom) panel 
shows the ratio of the measured cross sections to the FONLL calculation for 
charm~(bottom) production. The shaded area shows the uncertainty in the 
FONLL prediction. The larger mass makes this uncertainty smaller in the 
case of bottom quarks. These calculations agree with the data for bottom 
production. The same is true for charm within the theoretical uncertainty 
with a ratio of data/FONLL of $\sim$2. A similar tendency was obtained at 
the Tevatron~\cite{mateo1,cdf1}.

The electron spectrum from bottom shown in Fig.~\ref{fig3} is integrated 
from $p_{\mathrm{T}}=3$~to~5~GeV/$c$ and gives $4.8 ^{+1.8}_{-1.6}({\rm 
stat}) {}^{+1.9}_{-1.8}({\rm sys})$nb. This spectrum is then extrapolated 
to $p_{\mathrm{T}}=0$ using the shape predicted by pQCD. {\sc pythia} with 
varying intrinsic $k_{\rm T}$~($1.5\!<\!k_{\rm T}\!<\!10$~GeV/$c$) and 
FONLL with varying factorization~($\mu_F$) and 
renormalization~($\mu_R$) 
scales~($0.5\!<\!\mu_{F,R}/\sqrt{m^2+p^2_{\rm T}}\!<\!2$) are used to 
evaluate the systematic uncertainty~(12\%) to this extrapolation. The 
extrapolation results in a bottom cross section at mid-rapidity of 
$d\sigma_{b\bar{b}}/dy \mid_{y=0} = 0.92 ^{+0.34}_{-0.31}({\rm stat}) 
{}^{+0.39}_{-0.36}({\rm sys})\mu {\rm b}$, using a $b \rightarrow e$ total 
branching ratio of $10 \pm 1 \%$, calculated using the heavy flavor hadron 
ratios described above. Using {\sc hvqmnr}~\cite{hvq1} with 
{\sc cteq5m}~\cite{pdf1} 
parton distribution functions (PDF's) to integrate over rapidity, the 
total bottom cross section is 
determined to be $\sigma_{b\bar{b}}= 3.2 ^{+1.2}_{-1.1}({\rm stat}) 
{}^{+1.4}_{-1.3}({\rm sys}) \mu {\rm b}$. Various PDF's and bottom mass 
values are used to evaluate the systematic uncertainty~(8\%) of the 
rapidity extrapolation. This result is consistent with our result from the 
dielectron spectrum, which gave $\sigma_{b\bar{b}}= 3.9 \pm 2.5({\rm stat}) 
^{+3}_{-2}({\rm sys}) \mu {\rm b}$~\cite{dipp}. FONLL predicts 
$\sigma_{b\bar{b}}=1.87^{+0.99}_{-0.67} \mu {\rm b}$, in agreement with 
both these experimental results.

The fraction of bottom in heavy flavor electrons is found to be larger 
than 0.33 with 90\% confidence level at $p_{\mathrm{T}}>5$~GeV/$c$. 
Furthermore, the assumption of no bottom suppression directly 
leads to a lower limit on the nuclear modification factor of single 
electrons, $R_{\rm AA}$, of greater than 0.33 with the same confidence 
level. However, according to our measurements, $R_{\rm AA}$ is 
$\sim0.25\pm0.05({\rm stat})\pm0.05({\rm sys})$ at 
$5<p_{\mathrm{T}}<6$~GeV/$c$~\cite{hq2} in the 0-10\% central Au+Au 
collisions. At the same time the current level of uncertainty in the 
measurement precludes us from placing significant limits on the possible 
energy loss of bottom quarks.

In conclusion, the ratio of the yield of electrons from bottom to that 
from charm has been measured in $p$+$p$ collisions at $\sqrt{s}=$200~GeV. 
The ratio provides the first measurement of the spectrum of electrons from 
bottom at RHIC.  FONLL calculations~\cite{mateo2} agree with this result,  
which provides an important baseline for the study of heavy quark 
production in the hot and dense matter created in Au+Au collisions.

We thank the staff of the Collider-Accelerator and
Physics Departments at BNL for their vital contributions.
We acknowledge support from the Office of Nuclear Physics 
in DOE Office of Science, NSF and a sponsored research grant 
from Renaissance Technologies (U.S.A.),
MEXT and JSPS (Japan),
CNPq and FAPESP (Brazil),
NSFC (China),
MSMT (Czech Republic),
IN2P3/CNRS, and CEA (France),
BMBF, DAAD, and AvH (Germany),
OTKA (Hungary),
DAE (India),
ISF (Israel),
KRF and KOSEF (Korea),
MES, RAS, and FAAE (Russia),
VR and KAW (Sweden),
U.S. CRDF for the FSU,
US-Hungary Fulbright,
and US-Israel BSF.




\begin{references}

\bibitem{mateo1} D.~Acosta {\it et al.}, 
Phys. Rev. D {\bf 71}, 032001 (2005).

\bibitem{cdf1} D.~Acosta {\it et al.}, 
Phys. Rev. Lett. {\bf 91}, 241804 (2003).

\bibitem{hq1} S.~S.~Adler {\it et al.}, 
Phys. Rev. Lett. {\bf 97}, 252002 (2006).

\bibitem{hq2} S.~S.~Adler {\it et al.}, 
Phys. Rev. Lett. {\bf 98}, 172301 (2007).

\bibitem{jet1} M. Gyulassy and M. Plumer, 
Phys. Lett. {\bf B243},  432 (1990).

\bibitem{hq3} Y.~L.~Dokshitzer and D.~E.~Kharzeev, 
Phys. Lett. {\bf B519}, 199 (2001).

\bibitem{hq5} H.~van~Hees {\it et al.}, 
Phys. Rev. Lett. {\bf 73}, 192301 (2008).

\bibitem{ph1} K.~Adcox {\it et al.}, 
Nucl. Instrum. Methods Phys. Res. Sect.  {\bf A499},  469 (2003).

\bibitem{ph2} S.~S.~Adler {\it et al.}, 
Phys. Rev. C {\bf 75},  024909 (2007).

\bibitem{morino} Y.~Morino, arXiv:0903.3504 [nucl-ex].

\bibitem{ua1} C.~Albajar {\it et al.}, 
Phys. Lett. {\bf B186},  237 (1987).

\bibitem{mot1} T.~Sjostrand, Comput. Phys. Commum. {\bf 82},  74 (1994).

\bibitem{mot2} D.~Lange, Nucl. Instrum. Methods {\bf A462}, 152 (2001). 

\bibitem{bib:cratio1} D.~Acosta {\it et al.}, 
Phys. Rev. Lett. {\bf 91}, 241804 (2003).

\bibitem{bib:cratio2} L.~Gladilin, hep-ex/9912064.

\bibitem{bib:cratio3} G.~Alves {\it et al.}, 
Phys. Rev. Lett {\bf 77}, 2388 (1996).

\bibitem{bib:PDG} Particle Data Group 2007.

\bibitem{mateo2} M.~Cacciari {\it et al.}, 
Phys. Rev. Lett {\bf 95}, 122001 (2005); private communication

\bibitem{hvq1} M.~L.~Mangano {\it et al.},
  Nucl. Phys.  {\bf B405}, 507 (1993).

\bibitem{pdf1} H.~L.~Lai {\it et al.}, 
Eur. Phys. J. {\bf C12}, 375 (2000).

\bibitem{dipp} A.~Adare {\it et al.} 
Phys. Lett. {\bf B670}, 313 (2009).

\end{references}
\end{document}